\documentclass[traditabstract]{aa}

\usepackage{graphicx,amsmath}
\usepackage{amssymb}
\usepackage{color}
\usepackage{natbib}             
\usepackage{url}
\usepackage{multirow}
\usepackage{threeparttable}
\bibpunct{(}{)}{;}{a}{}{,} 

\begin{document}

   \title{Reconnaissance of the TRAPPIST-1 exoplanet system in the Lyman-$\alpha$ line}
                                   
   \author{
   V.~Bourrier\inst{1},                 
   D.~Ehrenreich\inst{1},               
   P.J.~Wheatley\inst{2},               
   E.~Bolmont\inst{3},                  
   M.~Gillon\inst{4},                   
        J.~de Wit\inst{5},                      
        A.J~Burgasser\inst{6},          
        E.~Jehin\inst{4},                       
        D.~Queloz\inst{7},\inst{1},                     
        A.H.M.J.~Triaud\inst{8}         
        }

\authorrunning{V.~Bourrier et al.}
\titlerunning{Observing TRAPPIST-1 at Ly-$\alpha$}

\offprints{V.B. (\email{vincent.bourrier@unige.ch})}

\institute{
Observatoire de l'Universit\'e de Gen\`eve, 51 chemin des Maillettes, 1290 Sauverny, Switzerland
\and 
Department of Physics, University of Warwick, Coventry CV4 7AL, UK
\and
Laboratoire AIM Paris-Saclay, CEA/DRF - CNRS - Univ. Paris Diderot - IRFU/SAp, Centre de Saclay, F- 91191 Gif-sur-Yvette Cedex, France
\and
Institut d’Astrophysique et de G\'eophysique, Universit\'e de Li\`ege, All\'ee du 6 Aout 19C, 4000 Li\`ege, Belgium
\and
Department of Earth, Atmospheric and Planetary Sciences, Massachusetts Institute of Technology, 77 Massachusetts Avenue, Cambridge, MA 02139, USA
\and
Center for Astrophysics and Space Science, University of California San Diego, La Jolla, CA 92093, USA
\and
Cavendish Laboratory, J J Thomson Avenue, Cambridge, CB3 0HE, UK
\and
Institute of Astronomy, Madingley Road, Cambridge CB3 0HA, UK
}
   
   \date{} 
 
  \abstract
{The TRAPPIST-1 system offers the opportunity to characterize terrestrial, potentially habitable planets orbiting a nearby ultracool dwarf star. We performed a four-orbit reconnaissance with the Space Telescope Imaging Spectrograph onboard the Hubble Space Telescope to study the stellar emission at Lyman-$\alpha$, to assess the presence of hydrogen exospheres around the two inner planets, and to determine their UV irradiation. We detect the Lyman-$\alpha$ line of TRAPPIST-1, making it the coldest exoplanet host star for which this line has been measured. We reconstruct the intrinsic line profile, showing that it lacks broad wings and is much fainter than expected from the stellar X-ray emission. TRAPPIST-1 has a similar X-ray emission as Proxima Cen but a much lower Ly-$\alpha$ emission. This suggests that TRAPPIST-1 chromosphere is only moderately active compared to its transition region and corona. We estimated the atmospheric mass loss rates for all planets, and found that despite a moderate extreme UV emission the total XUV irradiation could be strong enough to strip the atmospheres of the inner planets in a few billions years. We detect marginal flux decreases at the times of TRAPPIST-1b and c transits, which might originate from stellar activity, but could also hint at the presence of extended hydrogen exospheres. Understanding the origin of these Lyman-$\alpha$ variations will be crucial in assessing the atmospheric stability and potential habitability of the TRAPPIST-1 planets.}


\maketitle

\section{Introduction}
\label{intro} 

For close-in planets, heating by the stellar energy can lead to the expansion of the upper atmospheric layers and their eventual escape. Transit spectroscopy in the UV is a powerful way to probe these extended atmospheres, in particular in the stellar Lyman-$\alpha$ (Ly-$\alpha$) line of neutral hydrogen (e.g., \citealt{VM2003}, \citealt{Lecav2012}). To date, the warm Neptune GJ436b is the lowest-mass planet found evaporating (\citealt{Ehrenreich2015}), with transit absorption depths up to 60\% in the Ly-$\alpha$ line. The formation of such an extended exosphere is due in great part to the low mass of GJ436b and the gentle irradiation from its M-dwarf host (\citealt{Bourrier2015_GJ436}, \citealt{Bourrier2016}), suggesting that small planets around cold stars can produce very deep UV transit signatures. For telluric planets, large amounts of hydrogen in their upper atmosphere could indicate the presence of a steam envelope being photo-dissociated, and replenished by evaporating water oceans (\citealt{Jura2004}). Few attempts have been made, however, to detect atmospheric escape from Earth-sized planets, and Ly-$\alpha$ transit observations of the super Earth 55 Cnc e (\citealt{Ehrenreich2012}) and HD\,97658 b (\citealt{Bourrier2016_HD976}) showed no evidence for hydrogen exospheres. \\
The nearby ultracool dwarf TRAPPIST-1 was found to harbor seven short-period, transiting, Earth-sized planets (\citealt{Gillon2016,Gillon2017}). TRAPPIST-1\,e, f, and g are within the habitable zone, and all planets have equilibrium temperatures low enough to make liquid water possible on their surface. The combined transmission spectrum of the inner planets, b and c, in the near-infrared (\citealt{deWit2016}) revealed no significant features, which rules out cloud-free hydrogen-dominated atmospheres but still allows for denser envelopes, such as a cloud-free water-vapor atmosphere. The high level of X-rays to ultraviolet (XUV, in 5--912\,\AA) irradiation (\citealt{Wheatley2016}) and the sustained activity of M dwarfs over approximately one billion years could however impact the putative atmospheres of the TRAPPIST-1 planets and their habitability (\citealt{Miguel2015}, \citealt{Bolmont2016}).  \\
The system proximity (12\,pc), its large planet-to-star radii ratios, and high systemic velocity (-56\,km\,s$^{-1}$) make it an interesting target for Ly-$\alpha$ transit spectroscopy. Little is known about the FUV emission of such cold stars, and we present in this Letter a reconnaissance study of TRAPPIST-1 with the Hubble Space Telescope (HST). Our objectives were to try and measure the stellar Ly-$\alpha$ emission, to estimate the extreme UV (EUV, in 100--912\,\AA) irradiative environment of the planets, and to search for signatures of planetary hydrogen escape.\\

\section{The Ly-$\alpha$ line of TRAPPIST-1}
\label{sec:Ly_line}

\begin{figure}     
\includegraphics[trim=0.7cm 5.5cm 1.cm 8cm,clip=true,width=\columnwidth]{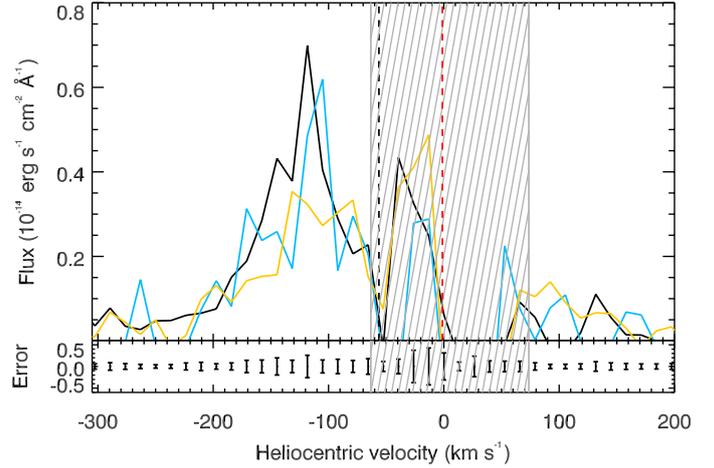}
\caption[]{Ly-$\alpha$ line spectrum of TRAPPIST-1, plotted as a function of Doppler velocity in the heliocentric rest frame. The black line has been averaged over the out-of-transit observations, while the blue and orange spectra were obtained at the times of TRAPPIST-1b and c transits, respectively. The dashed black line indicates the velocity of the star, and the dashed red line the velocity of the ISM/LIC cloud. The red wing of the Ly-$\alpha$ line cannot be observed from Earth because of ISM absorption along the line of sight, and contamination from geocoronal emission (hatched region).}
\label{fig:spectra}
\end{figure}

We used the Space Telescope Imaging Spectrograph (STIS) instrument onboard the HST to observe TRAPPIST-1 at FUV wavelengths (G140M grating at 1222\,\AA). Four HST orbits were obtained in 2016 in the frame of the Mid-Cycle Program 14493 (PI: Bourrier), with exposure time of approximately 35\,min. We observed TRAPPIST-1 during two consecutive HST orbits on 26 September, outside of the two inner planets' transits, to search for the stellar Ly-$\alpha$ emission and obtain a reference for the out-of-transit flux. We observed the star during the transit of TRAPPIST-1b on September 30th (one HST orbit) and approximately 1.7\,h after the transit of TRAPPIST-1c on November 23rd (one HST orbit). The data were reduced with the CALSTIS pipeline. In the region of the Ly-$\alpha$ line, the background is dominated by Earth geocoronal airglow emission. The airglow varies in strength and position with the epoch of observation, and its correction by the pipeline can yield spurious flux values when the airglow is much stronger than the stellar line. We thus excluded from our analysis the contaminated ranges [-60 ; 74]\,km\,s$^{-1}$ (out-of-transits), [-60 ; 49]\,km\,s$^{-1}$ (TRAPPIST-1b), and [-60 ; 54]\,km\,s$^{-1}$ (TRAPPIST-1c), defined in the heliocentric rest frame (Fig.~\ref{fig:spectra}). No breathing effect was detected in the time-tagged data (\citealt{Bourrier2016_HD976}), most likely because it is dominated by the photon noise from the very faint Ly-$\alpha$ line.

Stellar Ly-$\alpha$ lines are difficult to observe because of absorption by the interstellar medium (ISM) between the star and the Earth. However, the proximity of the TRAPPIST-1 system (12\,pc) limits the depth of this absorption, and the high systemic velocity of the star (-56.3\,km\,s$^{-1}$; \citealt{Reiners2015}) shifts the stellar line blueward of the spectrum. This allowed us to measure the blue wing and the core of TRAPPIST-1 Ly-$\alpha$ line, freed up from both ISM and airglow contamination (Fig.~\ref{fig:spectra}). The stellar line is spectrally resolved with STIS, and  measured with similar flux levels in our four epochs of observations. To our knowledge, TRAPPIST-1 is the coldest exoplanet host star from which UV emission has been detected, and the coldest star with a resolved observation of the Lyman-$\alpha$ line. We discuss the importance of this detection in Sect.~\ref{sec:EUV_irrad}.

\begin{figure}     
\includegraphics[trim=0.cm 5.5cm 0cm 10.7cm,clip=true,width=\columnwidth]{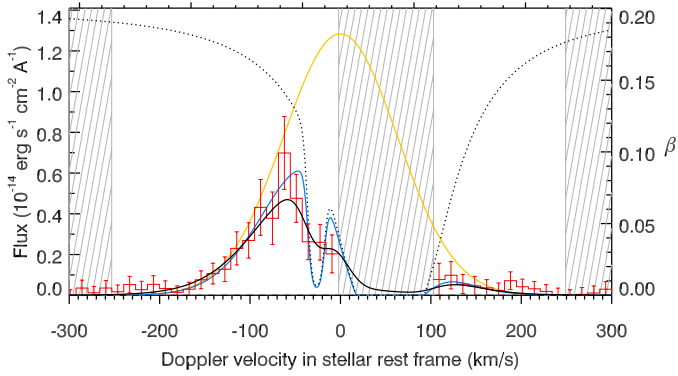}
\caption[]{Intrinsic Ly-$\alpha$ line profile of TRAPPIST-1 (orange line). The solid blue line shows the Ly-$\alpha$ line profile after absorption by the ISM hydrogen and deuterium (dotted black line). The solid black line shows the line profile convolved with STIS LSF and compared to the observations shown as a red histogram. Hatched regions were excluded from the fit. The right axis shows the ratio $\beta$ between radiation pressure and stellar gravity.}
\label{fig:theo_spectrum}
\end{figure}

We reconstructed the intrinsic Ly-$\alpha$ profile using the method detailed in \citet{Bourrier2015_GJ436}. A model profile of the stellar Ly-$\alpha$ line is absorbed by the interstellar hydrogen and deuterium, and convolved by STIS line spread function (LSF). A Metropolis-Hasting Markov chain Monte Carlo algorithm is used to find the best fit between the model spectrum and the observed spectrum, which are compared over the range shown in Fig.~\ref{fig:theo_spectrum}. The observed spectrum was calculated as the mean of all spectra, excluding the ranges contaminated by airglow emission or displaying possible flux variations (Sec.~\ref{sec:H_exos}). The heliocentric velocity of the star was fixed to its aforementioned value. The LISM Kinematic Calculator\footnote{\mbox{\url{http://sredfield.web.wesleyan.edu/}}} predicts that the line of sight (LOS) toward TRAPPIST-1 crosses the Local Interstellar Cloud (LIC). Since the fit is mostly sensitive to the hydrogen column density in the cloud, we fixed its heliocentric radial velocity, temperature, and turbulent velocity to the values of the LIC (-1.25\,km\,s$^{-1}$, 7500\,K, 1.62\,km\,s$^{-1}$, respectively; \citealt{Redfield_Linsky2008}). Trying out different profiles for the intrinsic stellar Ly-$\alpha$ line, we found it was best represented as a single-peaked Gaussian profile, displayed in Fig.~\ref{fig:theo_spectrum}. The best-fit hydrogen column density for the ISM is log$_{10}$\,$N$(H\,{\sc i}) = 18.3$\pm$0.2, similar to other LOS at a distance of $\sim$12\,pc (Fig. 14 in \citealt{Wood2005}).  \\

\section{High-energy emission from an ultracool dwarf}
\label{sec:EUV_irrad} 

The X-ray emission of TRAPPIST-1 was recently studied by \citet{Wheatley2016}, who found that the star is a relatively strong and variable coronal X-ray source with an X-ray luminosity similar to that of the quiet Sun, despite a much lower bolometric luminosity. We report the weighted average of their APEC- and CEMEKL-derived fluxes in 5--100\,\AA\, in Table~\ref{tab:XEUV_flux}. We searched the GALEX data\footnote{http://galex.stsci.edu/galexview/} for near-UV (NUV) or far-UV (FUV) observations of TRAPPIST-1, but the closest star with a detection is 57'' away. The stellar EUV emission would, in any case, be absorbed by the ISM, but it can be estimated through indirect methods. We used the \citet{Linsky2014} relation for M dwarfs, based on our estimate for the intrinsic Ly-$\alpha$ flux, to calculate the flux in 100--912\,\AA\, (Table~\ref{tab:XEUV_flux}). We find a much lower EUV flux than derived by \citet{Wheatley2016} from the X-ray flux and \citet{Chadney2015} empirical relation, possibly due to the very late type of TRAPPIST-1.

Little is known about the UV emission of low-mass, cool stars such as TRAPPIST-1 (Table~\ref{tab:Mdwarfs}). Because of their low temperature, the photosphere of M dwarfs contributes little to the UV continuum, compared to bright chromospheric and transition region emission lines. Ly-$\alpha$ in particular is the brightest UV line of low-mass stars (e.g., \citealt{France2013}). Assuming the \citet{Linsky2014} relations are relevant to such a cold star, the Ly-$\alpha$ flux of TRAPPIST-1 is indeed as large as its entire EUV flux. In Table~\ref{tab:Mdwarfs}, we compare TRAPPIST-1 with the coldest M dwarf hosts with a measured Ly-$\alpha$ flux. Ly-$\alpha$ emission decreases with stellar temperature (\citealt{Linsky2013}). TRAPPIST-1 is more than 400\,K colder than those stars, hence it is not surprising that its Ly-$\alpha$ flux is more than twice as small. Interestingly, TRAPPIST-1 has a similar X-ray emission as Proxima Cen (0.136 and 0.142\,erg\,s$^{-1}$\,cm$^{-2}$ at 1\,au, respectively) but a much lower Ly-$\alpha$ emission. This might be an indication of a lower chromospheric activity for TRAPPIST-1 compared to its corona, and is consistent with the decrease in the ratio of Ly-$\alpha$ to X-ray emission noted for later-type stars by \citet{Linsky2013}. Furthermore, XUV activity as a function of age has been shown to remain constant for M stars younger than approximately 300\,Myr (\citealt{Shkolnik2014}). After this saturation phase, chromospheric activity and coronal emission steadily decrease due to the spin down of the star, but the decrease in X-ray emission is steeper than in the NUV and FUV (\citealt{Stelzer2013}). The fact that TRAPPIST-1 emits nearly three times less flux at Ly-$\alpha$ than in the X-ray would thus suggest it is still relatively young (age is constrained to be more than 500\,Myr, \citealt{Filippazzo2015}).\\

\begin{table}
\centering
\caption{XUV irradiation of TRAPPIST-1 planets.}
\label{tab:XEUV_flux}
\begin{threeparttable}
\begin{tabular}{lccccc}
\hline
\hline
\noalign{\smallskip}    
 & F$_\mathrm{X}$& F$_\mathrm{EUV}$ & F$_{Ly\alpha}$ & $\Gamma_{\mathrm{ion}}$  & $\dot{\mathrm{M}}^{\mathrm{tot}}$ \\  
 & (5-100\,\AA)   &   (100-912\,\AA) &    &   &   \\     
\noalign{\smallskip}
\hline
\noalign{\smallskip}
T-1b & 1102$\stackrel{+473}{_{-156}}$  & 400$\stackrel{+344}{_{-264}}$  & 414$\stackrel{+102}{_{-142}}$  &  7.6$\stackrel{+8.1}{_{-5.5}}$ & 46$\stackrel{+25}{_{-13}}$  \\
T-1c & 588$\stackrel{+ 252}{_{-83}}$ &  213$\stackrel{+184}{_{-141}}$&221$\stackrel{+54}{_{-76}}$ &4.1$\stackrel{+4.3}{_{-2.9}}$ &14$\stackrel{+8}{_{-4}}$ \\
T-1d & 296$\stackrel{+127}{_{-42}}$ &107$\stackrel{+92}{_{-71}}$ &111$\stackrel{+27}{_{-38}}$ &2.0$\stackrel{+2.2}{_{-1.5}}$ &9$\stackrel{+5}{_{-3}}$ \\
T-1e & 171$\stackrel{+74}{_{-24}}$ & 62$\stackrel{+54}{_{-41}}$ &64$\stackrel{+16}{_{-22}}$ &1.2$\stackrel{+1.3}{_{-0.9}}$ &6$\stackrel{+3}{_{-2}}$ \\
T-1f & 99$\stackrel{+42}{_{-14}}$ &36$\stackrel{+31}{_{-24}}$ &37$\stackrel{+9}{_{-13}}$ &0.7$\stackrel{+0.7}{_{-0.5}}$ &5$\stackrel{+3}{_{-1}}$ \\
T-1g & 67$\stackrel{+29}{_{-10}}$ &24$\stackrel{+21}{_{-16}}$ &25$\stackrel{+6}{_{-9}}$ &0.5$\stackrel{+0.5}{_{-0.3}}$ &2$\stackrel{+1}{_{-1}}$ \\T-1h & 34$\stackrel{+15}{_{-5}}$ &12$\stackrel{+11}{_{-8}}$ &13$\stackrel{+3}{_{-4}}$  &0.2$\stackrel{+0.3}{_{-0.2}}$  &-  \\
\noalign{\smallskip}
\hline  
\hline
\end{tabular}
  \begin{tablenotes}[para,flushleft]
  Note: Fluxes in erg\,s$^{-1}$\,cm$^{-2}$, hydrogen photoionization rate $\Gamma_{\mathrm{ion}}$ in x10$^{-6}$\,s$^{-1}$, mass loss rate $\dot{\mathrm{M}}^{\mathrm{tot}}$ in 10$^{6}$\,g\,s$^{-1}$. X-ray flux is derived from \citet{Wheatley2016}. The mass of TRAPPIST-1\,h is unconstrained, and we did not estimate its mass-loss rate. Uncertainties on the EUV flux were estimated by accounting for both the uncertainty on the Ly-$\alpha$ flux and the \citet{Linsky2014} relation. \\
  \end{tablenotes}
  \end{threeparttable}
\end{table}

\begin{table}
\centering
\caption{Coldest M dwarf hosts with measured Ly-$\alpha$ flux}
\label{tab:Mdwarfs}
\begin{threeparttable}
\begin{tabular}{lccc}
\hline
\hline
\noalign{\smallskip}    
Star & Type & T$_\mathrm{eff}$ & F$_{Ly\alpha}$         \\      
\noalign{\smallskip}
     &    &  (K)  &  (erg\,s$^{-1}$\,cm$^{-2}$)  \\      
\noalign{\smallskip}
\hline
\noalign{\smallskip}
TRAPPIST-1  & M8  &  2550$\pm55^{[1]}$    & 0.05$\stackrel{+0.01}{_{-0.02}}$                                    \\ 
GJ\,1214   & M4.5  &  2935$\pm100^{[2]}$    &  0.12$\stackrel{+0.13}{_{-0.05}}^{[3]}$                                    \\ 
Proxima Cen  & M5.5  &  3050$\pm100^{[4]}$    & 0.30$\pm0.06^{[5]}$                                         \\
GJ\,876  & M5  &  3062$\pm130^{[2]}$    & 0.37$\pm0.04^{[3]}$                                      \\  
\noalign{\smallskip}
\hline
\hline
\end{tabular}
  \begin{tablenotes}[para,flushleft]
  Note: Fluxes are given at 1\,au from the stars. References: $^{[1]}$\citealt{Gillon2016}; $^{[2]}$\citealt{Loyd2016}; $^{[3]}$\citealt{Youngblood2016}; $^{[4]}$\citealt{Anglada2016}; $^{[5]}$\citealt{Wood2005}.  \\
  \end{tablenotes}
  \end{threeparttable}
\end{table}

\section{Mass loss from the TRAPPIST-1 planets}
\label{sec:mass_loss}

While the TRAPPIST-1 planets are unlikely to be surrounded by hydrogen-dominated atmospheres (\citealt{deWit2016}), they may still harbor a substantial amount of water (\citealt{Gillon2017}). The stellar XUV irradiation and internal tidal heating could be sufficient for water oceans to form a supercritical steam atmosphere that could be photodissociated and sustain a significant hydrogen escape (\citealt{Jura2004}). Using the derived XUV flux, we calculated the mass-loss rate from all TRAPPIST-1 planets in the energy-limited regime (\citealt{Lecav2007}), updating the estimates made by \citet{Bolmont2016} and \citet{Wheatley2016}:
\begin{equation}
\label{eq:H_esc_rate}
\dot{\mathrm{M}}^{\mathrm{tot}}= \eta \, (\frac{R_\mathrm{XUV}}{R_\mathrm{p}})^2 \, \frac{3 \, F_\mathrm{XUV}(\mathrm{sma})}{4 \, G \, \rho_\mathrm{p} \, K_{tide}},  
\end{equation}
with $\eta$ representing the heating efficiency set to 1\% (\citealt{Ehrenreich2015}, \citealt{Salz2016a}). $(\frac{R_\mathrm{XUV}}{R_\mathrm{p}})^2$ accounts for the increased cross-sectional area of planets to EUV radiation, while $K_\mathrm{tide}$ accounts for the contribution of tidal forces to the potential energy (\citealt{Erkaev2007}). Both are set to unity for these cool and small planets. The mean density of the planets was calculated using preliminary masses derived from transit timing variations (\citealt{Gillon2017}). The derived escape rates are given in Table~\ref{tab:XEUV_flux}. Earth-like atmospheres ($\sim$5$\times$10$^{21}$\,g) and oceans ($\sim$1.4$\times$10$^{24}$\,g) would take between 5 and 22\,Gy to be stripped from TRAPPIST-1d to g, but only approximately 1 and 3\,Gy for TRAPPIST-1b and c. While this result strongly depends on the heating efficiency and neglects the complex physics behind atmospheric escape (Bolmont et al. 2017, in prep), it suggests nonetheless that Earth-like TRAPPIST-1b and c would have to be still young - or to have an initial composition richer in water than Earth - to have retained a significant water content.

\section{Hydrogen exospheres around TRAPPIST-1b/c}
\label{sec:H_exos} 

\begin{figure}     
\includegraphics[trim=1.cm 3.1cm 2cm 9cm,clip=true,width=\columnwidth]{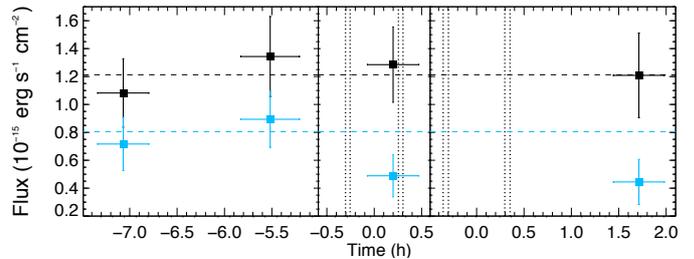}
\caption[]{Ly-$\alpha$ line flux integrated over [-95 ; -55]\,km\,s$^{-1}$ (blue points) and the complementary domain in [-210 ; 210]\,km\,s$^{-1}$, airglow excluded (black points). The flux is normalized by the reference measurements in the left panel, and plotted as a function of time relative to the transit of TRAPPIST-1b (left and middle panels) and TRAPPIST-1c (right panel).}
\label{fig:lc}
\end{figure}

Using the out-of-transit Ly-$\alpha$ line as a reference, we identified marginal flux decreases (Fig.~\ref{fig:lc}) during the transit of TRAPPIST-1b (40$\pm$21\% in [-95 ; -55]\,km\,s$^{-1}$) and after the transit of TRAPPIST-1c (41$\pm$18\% in [-135 ; -40]\,km\,s$^{-1}$). Since the star has a variable corona (\citealt{Wheatley2016}), this might be an indication of a similarly variable chromosphere. Alternatively, and given that the Ly-$\alpha$ line is stable over time outside of the above ranges (Fig.~\ref{fig:lc}), this might hint at the presence of hydrogen exospheres around the two inner planets. The measured variations are too large to be caused by the planetary disks (\citealt{Llama2016}) and would correspond to opaque exospheric disks extending up to approximately seven times the planets' radii. We note that the putative exosphere of TRAPPIST-1c would absorb the stellar Ly-$\alpha$ line up to two hours after the planet transit, at larger velocities than the signal from TRAPPIST-1b. This suggests that the exosphere of planet c would extend into a cometary tail, with the coma around the planet likely producing a much deeper absorption at the time of the transit.\\
The velocity field of these exospheres would depend on the planet orbital velocity and the relative strength of radiation pressure and stellar gravity (\citealt{Bourrier2015_GJ436}). The faint Ly-$\alpha$ line of TRAPPIST-1 yields a radiation pressure that can only compensate for up to 20\% of the star gravity (Fig.~\ref{fig:theo_spectrum}). This is reminiscent of the giant hydrogen exosphere of GJ\,436 b (\citealt{Ehrenreich2015}), slowly diffusing around the planet owing to the low radiation pressure from its M dwarf host ($\beta<$70\%). However, using the EVE code (\citealt{Bourrier_lecav2013}) to simulate atmospheric escape from TRAPPIST-1b and c, we found that the orbital velocities of the planets ($\sim$80 and 70\,km\,s$^{-1}$, respectively) and the low radiation pressure could impart maximum Doppler velocities below $\sim$40\,km\,s$^{-1}$ to hydrogen atoms at the time of the observations. If the absorption signatures from TRAPPIST1-b and -c are confirmed, this means that their exospheres are subjected to additional dynamical mechanisms (e.g., stellar wind interactions or a high-speed planetary outflow). Finally, we used the XUV spectrum derived in Sect.~\ref{sec:Ly_line} to calculate the photoionization rate of neutral hydrogen at the semi-major axis of TRAPPIST-1b and c. We found low values that correspond for each planet to lifetimes of approximately 40 and 70\,hours (Table~\ref{tab:XEUV_flux}). This ionization field, lower than for GJ\,436 b (\citealt{Bourrier2016}), would further help the formation of giant hydrogen exospheres around the inner TRAPPIST-1 planets. \\

\section{Conclusion}
\label{sec:conclu}

The TRAPPIST-1 system is a compelling target for atmospheric characterization, with seven temperate terrestrial planets. We detect the Ly-$\alpha$ of its ultracool dwarf star, and derive the EUV irradiation of the planets. The Ly-$\alpha$ line contributes as much as the entire EUV spectrum, and combined with an X-ray irradiation more than twice as strong, it could significantly impact the chemical and physical balance of the planets' atmospheres. In particular, water vapor in the upper atmosphere could photo-dissociate and sustain an outflow of escaping hydrogen. The stellar Ly-$\alpha$ line is bright enough to perform transit spectroscopy, and we detect marginal flux decreases in localized, high-velocity ranges during the transit of planet b, and shortly after the transit of planet c. This could hint at the presence of extended hydrogen exospheres around the two inner planets, and suggest that atmospheric escape might play an active role in the evolution of all TRAPPIST-1 planets. Even if the measured variations do not have a planetary origin and arise from activity in the stellar Ly-$\alpha$ line, this would have important consequences for the stability of the planetary atmospheres and the production of biomarkers. More observations will be required to characterize the high-energy environment of the TRAPPIST-1 system, and its effect on the planets' atmospheres.


\begin{acknowledgements}
We thank the referee for their efficient reading of our paper. This work is based on observations made with the NASA/ESA Hubble Space Telescope, for which support was provided by NASA through a grant from the Space Telescope Science Institute. This work has been carried out in the frame of the National Centre for Competence in Research ``PlanetS'' supported by the Swiss National Science Foundation (SNSF), and was funded in part by the European Research Council (ERC) under the FP/2007-2013 ERC grant 336480. V.B. and D.E. acknowledges the financial support of the SNSF. M.G. and E.J. are Research Associate at the Belgian Fund for Scientific Research (F.R.S-FNRS). E. B. acknowledges funding by the European Research Council through ERC grant SPIRE 647383. PW is supported by  a Science and Technology Facilities Council (STFC) consolidated grant (ST/L000733/1).
\end{acknowledgements}

\bibliographystyle{aa} 
\bibliography{biblio} 

\end{document}